
\magnification=1200
\def\wc{\hangindent=4em \hangafter=1 \noindent}
\baselineskip 18pt
\parskip 3pt
\null

\headline={\ifnum\pageno=1\hfil\else\hfil\tenrm--\ \folio\ --\hfil\fi}
\footline={\hfil}

\centerline{\bf Gravitational Microlensing by the Globular Cluster Stars}
\vskip 0.5cm
\centerline{by}
\vskip 0.5cm
\centerline{Bohdan Paczy\'nski}
\vskip 0.5cm
\centerline{Princeton University Observatory, Princeton, NJ 08544-1001, USA}
\centerline{e-mail: bp@astro.princeton.edu}
\vskip 0.5cm
\centerline{\it Received July 28, 1994}
\vskip 0.5cm
\centerline{ABSTRACT}
\vskip 0.5cm
The microlensing of background stars by compact objects in globular
clusters is analyzed.  The main strength of the proposed search is the
direct relationship between the lens mass and the time scale of the
microlensing event.  The main problem is the low overall rate of
expected events which implies that a ground based search should last for
about a decade to generate a non trivial number of events, like a few
dozen.  A space borne experiment could increase the rate by a large
factor by increasing the number of stars which could be monitored thanks
to the much better seeing. The clusters M22 with the galactic bulge
background, and 47 Tuc with the SMC background, are the two examples of
possible targets.

\vskip 0.5cm
Key words: gravitational lensing - stars: clusters, low mass, brown dwarfs
\vskip 0.5cm

The effect of gravitational microlensing of the LMC, SMC, and the
galactic bulge stars by all intervening compact objects has been
proposed as a way to detect low mass, dark objects, like brown dwarfs
and even planets (Paczy\'nski 1986, 1991, Griest et al. 1991).
Recently, three groups reported the detection of the total of 17
microlensing event candidates: OGLE - 10 (Udalski et al. 1993,
1994a,b,c), MACHO - 5 (Alcock et al. 1993, 1994), and EROS - 2 (Auburg
et al. 1993).  The most thoroughly analyzed results were those of the
OGLE (Udalski et al. 1994b, Paczy\'nski et al. 1994), and these revealed
a very unpleasant reality: the determination of the masses of lensing
objects on the basis of the observed event time scales is very model
dependent.  In fact, there is so much uncertainty about the distribution
of ordinary stars, not to mention brown dwarfs and more exotic objects,
in our own galaxy that the mass determination is currently impossible.
The aim of this paper is to present the conditions under which the
relation between the time scale and the lens mass is very direct and
reasonably unique.

Consider rich globular clusters, like 47 Tuc or M22, seen against the
rich background of either SMC or the galactic bulge stars.  We propose
to consider the lensing events that are expected when any compact
globular cluster object passes in front of a distant star, which acts as
the source of light.  A typical velocity dispersion in a globular
cluster is $ \sim 10 ~ km ~ s^{-1} $, while a typical transverse
velocity of a cluster as a whole is $ \sim 100 ~ km  s^{-1} $.
Therefore, even if the lens cannot be directly detected its transverse
velocity is well approximated by that of a cluster as a whole.  The same
is true for the distance to the lens.  The lensed star is always
directly visible, and in most cases it is likely to be at the well known
distance, be it the SMC or the galactic bulge.  As a reasonable first
approximation one can take the distance to the source to be the same as
that of the average system it belongs to.  However, in principle its own
distance is measurable, as it is directly observed.

The geometry as outlined above offers a unique opportunity to derive the
lens mass when the event time scale is measured.  The combined
uncertainty due to the depth effects of the source system or the
velocity dispersion in the cluster may generate $ \sim 20 \% $
uncertainty in the lens mass.  The uncertainty with the currently
reported lensing events is more like a factor of 3 or even 10, as in
case of the LMC lensing events the lenses may be anywhere between the
galactic disk and the LMC bar (Gould et al. 1994, Sahu 1994).
Therefore, the lenses in globular clusters offer unprecedented accuracy
of the mass determination.

There is a very good reason to determine the mass function in globular
clusters, as there is observational indication that ``some, (and perhaps
all) clusters probably have very steep IMF's with the slope likely
exceeding 2.5 (Salpeter value 1.3)'' (Richer et al. 1991).  The
inference was based on the observed {\it luminosity} function and the
theoretical mass-luminosity relation.  The gravitational lensing offers
the only direct determination of the mass function in the globular
clusters which are seen against a rich background of reasonably bright
stars.

Let us determine the number of events expected in a whole cluster and
the relation between the event time scale and the lens mass following
Paczy\'nski (1991).  For simplicity the mass distribution in a globular
cluster is approximated here with a singular isothermal sphere truncated
at 1/3 of its tidal radius as estimated with the King model (King 1966).
 A singular model is wrong within the core radius, but this is of no
concern to us as no background stars can be detected within the core
radius.

The column mass density of a singular isothermal sphere is given as
$$
\Sigma \approx { \sigma ^2 \over 2Gr } , \eqno(1)
$$
where $r$ is the projected distance from the cluster center.
Let $ D_s $ and $ D_d $ be the distance from the observer to
the source and the lens (deflector), respectively.  The optical
depth to microlensing can be calculated to be
$$
\tau =
{ 4 \pi G \Sigma \over c^2 } D_d \left( 1 - { D_d \over D_s } \right) =
{ \sigma ^2 \over c^2 } { 2 \pi \over \varphi }
\left( 1 - { D_d \over D_s } \right) , \eqno(2)
$$
where $ \sigma $ is the one dimensional (radial) velocity dispersion,
and $ \varphi $ is the angular distance from the cluster center.
If the sources are at very large distance, $ D_d/D_s \ll 1 $, then
the correction term in the brackets is equal one, and the formula
is even simpler:
$$
\tau \approx 2.4 \times 10^{-5} \left( { \sigma \over 10 ~ km ~ s^{-1} }
\right) ^2 \left( { 1' \over \varphi } \right).  ~~~~~~~~~~~~
D_d/D_s \ll 1 . \eqno(3)
$$
This optical depth is much higher than typically expected towards
the galactic bulge or the SMC, i.e. within a few arcminutes of the
cluster center the lensing is likely to be dominated by the compact
objects associated with the cluster itself.  In any case the
``background'' due to the other galactic or LMC lenses can be readily
determined observationally.

Following Paczy\'nski (1991) we can also calculate the relation between
the time scale of the microlensing event, $ t_0 \equiv R_E/V $,
where $R_E$ is the Einstein ring radius, and V is the transverse
velocity in the observer-lens-source system.  The time scale
can be calculated as
$$
t_0 ~ [days] = 32.9 ~ \left( { M_d \over 0.1 ~ M_{\odot} } \right) ^{1/2}
\left( { 1 ~ kpc \over D_d } \right) ^{1/2}
\left( 1 - { D_d \over D_s } \right) ^{1/2}
\left( { 10 ~ mas ~ yr^{-1} \over \dot \varphi } \right), \eqno(4)
$$
where $ M_d $ is the lens (deflector) mass,
$ \dot \varphi $ is the observed relative proper motion of
the globular cluster with respect to the lensed star, and all other
symbols have their usual meaning.  As expected the time scale is
shorter when the lens mass is smaller, the distance to the lens is
larger, and its transverse angular velocity is larger.

Now we have to estimate the number of events that are expected
in one year of continuous monitoring of the cluster background.
The OGLE experience indicates (Udalski et al. 1992) that up to
400 stars can be measured per area $ (1')^2 $ in the sky while
the seeing is $ 1'' $ (FWHM).  Naturally, the better the seeing
the more stars can be measured in a dense stellar field.  We
adopt
$$
n \approx 400 \left( { 1'' \over FWHM } \right) ^2
{}~~~~~~~~~~ {\rm per} ~~ (1')^2 , \eqno(5)
$$
where $n$ is the number of measurable stars per $(1')^2$.
The number of measurable background stars that are microlensed
at any instant, i.e. which have their apparent brightness
increased by at least a factor 1.34, can be calculated as
$$
N = \int _{\varphi _1}^{\varphi _2} n \tau ~ 2 \pi \varphi ~ d \varphi
= 0.60 \left( { 1'' \over FWHM } \right) ^2
\left( { \varphi _2 - \varphi _1 \over 10' } \right)
\left( { \sigma \over 10 ~ km ~ s^{-1} } \right)  ^2
\left( 1 - { D_d \over D_s } \right) ,  \eqno(6)
$$
where $ \varphi _1 $ and $ \varphi _2 $ are the inner and the
outer radius of the annulus around the cluster within which
the search for microlensing is feasible.
The rate of events per year can be calculated as
$$
\Gamma ~ [yr^{-1}] = { 2N \over \pi } ~ { 1 ~ yr \over t_0 } =
$$
$$
4.2 \times
\left( { \sigma \over 10 ~ km ~ s^{-1} } \right)  ^2
\left( { 0.1 ~ M_{\odot} \over M_d } \right) ^{1/2}
\left( { \varphi _2 - \varphi _1 \over 10' } \right)
\left( { \dot \varphi \over 10 ~ mas ~ yr^{-1} } \right) \times
\eqno(7)
$$
$$
\left( { D_d \over 1 ~ kpc } \right) ^{1/2}
\left( 1 - { D_d \over D_s } \right) ^{1/2}
\left( { 1'' \over FWHM } \right) ^2 .
$$

The last two equations demonstrate the main problem with the idea
presented in this paper: the expected rate of events is very small
unless a substantial fraction of globular cluster mass is in very low
mass objects, or we can somehow reduce the FWHM of the seeing disk, for
example with a space borne experiment.  A traditional observing program
from the ground should extend for about a decade to generate non-trivial
 results.

There is yet another problem.  Our estimate of $ \sim 400 $ stars per
square arcminute at the 1'' seeing refers to the saturation limit: we
do not expect to be able to measure more stars than that number. In case
of M22 the background galactic bulge stars are bright enough that it
might be relatively easy to reach the saturation limit with  a 1 meter
class telescope.  The background of SMC stars in case of 47 Tuc is much
fainter (cf. Fig. 4 of Hesser {\it et al.} 1987), and it may take a 4
meter class telescope to carry out the observing program.

Naturally, the range of angular distances from the cluster center,
$ \varphi _1 \leq \varphi \leq \varphi _2 $, is limited by the
crowding of the clusters stars at $ \varphi _1 $, and the tidal radius
of the clusters at $ \varphi _2 $.

\vskip 0.5cm
{\bf Acknowledgements}.  It is a great pleasure to acknowledge
stimulating discussions with N. Reid and I. Thompson, and the hospitality
of A. Omont and many other astronomers at the Institut d'Astrophysique in
Paris where this paper has been written.  This project was
supported with the NSF grant AST92-16494.

\vskip 0.5cm
\centerline{REFERENCES}
\vskip 0.5cm

\wc{Alcock, Ch. {\it et al.} 1993, {\it Nature}, {\bf 365}, 621. \hfill}

\wc{Alcock, Ch. {\it et al.} 1994, {\it Astrophys. J. Letters}, submitted.
\hfill}

\wc{Aubourg, E. {\it et al.} 1993, {\it Nature}, {\bf 365}, 623.  \hfill}

\wc{Gould, A., Miralda-Escud\'e, J. and Bahcall, J. N. 1994, {\it Astrophys.
J}., {\bf 423}, L105.
\hfill}

\wc{Griest, K. {\it et al.} 1991, {\it Astrophys. J. Letters}, {\bf 372},
L79.  \hfill}

\wc{Hesser, J. E., Harris, W. E., Vandenberg, D. A., Allwright, J. W. B.,
Shott, P., and Stetson, P. B.  1987, {\it Publ. Astron. Soc. Pacific},
{\bf 99}, 739.  \hfill}

\wc{King, I. R. 1966, {\it Astron. J.}, {\bf 71}, 64.  \hfill}

\wc{Paczy\'nski, B. 1986, {\it Astrophys. J. Letters}., {\bf 304}, 1.  \hfill}

\wc{Paczy\'nski, B. 1991, {\it Astrophys. J. Letters}, {\bf 371}, L63.  \hfill}

\wc{Paczy\'nski, B., B, Stanek, K. Z., Udalski, A.. Szyma\'nski, M.,
Ka\l u\.zny, J., Kubiak, M., Mateo, M., \& Krzemi\'nski, W.
1994, {\it Astrophys. J. Letters}, submitted.  \hfill}

\wc{Richer, H. B., Fahlman, G. G., Buonanno, R., Pecc, F. F., Searle, L.,
and Thompson, I. B. 1991, {\it Astrophys. J.}, {\bf 381}, 147.  \hfill}

\wc{Sahu, K.  1994, {\it Nature}, ???, ...   \hfill}

\wc{Udalski, A., Szyma\'nski, M., Ka\l u\.zny, J., Kubiak, M., and Mateo, M.
1992, {\it Acta Astron}., {\bf 42}, 253.  \hfill}

\wc{Udalski, A., Szyma\'nski, M., Ka\l u\.zny, J., Kubiak, M., Krzemi\'nski,
W., Mateo, M., Preston, G. W., and Paczy\'nski, B.  1993, {\it Acta Astron}.,
{\bf 43}, 289.  \hfill}

\wc{Udalski, A., Szyma\'nski, M., Ka\l u\.zny, J., Kubiak, M., Mateo, M.,
and Krzemi\'nski, W. 1994a, {\it Astrophys. J. Letters}, {\bf 426}, L69.
\hfill}

\wc{Udalski, A., Szyma\'nski, M., Stanek, K. Z., Ka\l u\.zny, J.,
Kubiak, M., Mateo, M., Krzemi\'n- ski, W., Paczy\'nski, B., and Venkat, R.
1994b, {\it Acta Astron},. {\bf 44}, 165.  \hfill}

\wc{Udalski, A., Szyma\'nski, M., Mao, S., Di Stefano, R., Ka\l u\.zny, J.,
Kubiak, M., Mateo, M., and Krzemi\'n- ski, W.  1994c, {\it Astrophysical
J. Letters}, submitted  \hfill}

\vfill
\end
\bye